\documentclass[showpacs,aps,prl,reprint,superscriptaddress,preprintnumbers]{revtex4-1}

\usepackage{amsmath}
\usepackage{amssymb}
\usepackage{graphicx}
\usepackage{dcolumn}
\usepackage[dvipsnames]{xcolor}
\usepackage{color}
\usepackage{endnotes}
\usepackage{bm}
\usepackage{epsfig}
\usepackage{array}
\usepackage{ulem}

\newcommand{\etal}{{\it et al.}}

\newcommand{\MoGa}{{Mo$_{8}$Ga$_{41}$}}
\newcommand{\MoVGa}{{Mo$_7$VGa$_{41}$}}

\begin{document}



\title{Linear magnetoresistance with a universal energy scale \\ in a strong-coupling superconductor}

\author{W.~Zhang}
\author{Y.~J.~Hu}
\affiliation{Department of Physics, The Chinese University of Hong Kong, Shatin, Hong Kong} 
\author{C.~N.~Kuo}
\author{S.~T.~Kuo}
\affiliation{Department of Physics, National Cheng Kung University, Tainan 70101, Taiwan}
\author{Yue-Wen Fang}
\affiliation{NYU-ECNU Institute of Physics, NYU Shanghai, Shanghai 200062, China}
\affiliation{Department of Materials Science and Engineering, Kyoto University, Kyoto 606-8501, Japan}
\author{Kwing~To~Lai}
\author{X.~Y.~Liu}
\author{K.~Y.~Yip}
\affiliation{Department of Physics, The Chinese University of Hong Kong, Shatin, Hong Kong} 
\author{D. Sun} 
\author{F. F. Balakirev}
\affiliation{National High Magnetic Field Laboratory, Los Alamos National Laboratory, Los Alamos, New Mexico 87545, USA}
\author{C.~S.~Lue}
\affiliation{Department of Physics, National Cheng Kung University, Tainan 70101, Taiwan}
\author{Hanghui~Chen}
\email[]{hanghui.chen@nyu.edu}
\affiliation{NYU-ECNU Institute of Physics, NYU Shanghai, Shanghai 200062, China}
\affiliation{Department of Physics, New York University, New York 10003, USA}
\author{Swee~K.~Goh}
\email[]{skgoh@cuhk.edu.hk}
\affiliation{Department of Physics, The Chinese University of Hong Kong, Shatin, Hong Kong}

\date{August 10, 2020}

\begin{abstract}
 
The recent discovery of a nonsaturating linear magnetoresistance in several correlated electron systems near a quantum critical point 
has revealed an interesting interplay between the linear magnetoresistance and the zero-field linear-in-temperature resistivity. These studies suggest a possible role of quantum criticality on the observed linear magnetoresistance. Here, we report our discovery of a nonsaturating, linear magnetoresistance in Mo$_8$Ga$_{41}$, a nearly isotropic strong electron-phonon coupling superconductor with a linear-in-temperature resistivity from the transition temperature to $\sim$55~K. The growth of the resistivity in field is comparable to that in temperature, provided that both quantities are measured in the energy unit. Our datasets are remarkably similar to magnetoresistance data of the optimally doped La$_{2-x}$Sr$_x$CuO$_4$, despite the clearly different crystal and electronic structures, and the apparent absence of quantum critical physics in Mo$_8$Ga$_{41}$. A new empirical scaling formula is developed, which is able to capture the key features of the low-temperature magnetoresistance data of Mo$_8$Ga$_{41}$, as well as the data of La$_{2-x}$Sr$_x$CuO$_4$.

\end{abstract}
\maketitle


Recently, interesting cases of nonsaturating linear magnetoresistance (LMR) has been reported in several correlated electron systems, including CrAs under pressure, Ba(Fe$_{1/3}$Co$_{1/3}$Ni$_{1/3}$)$_2$As$_2$, La$_{2-x}$Ce$_x$CuO$_4$, La$_{2-x}$Sr$_x$CuO$_4$,  BaFe$_2$(As$_{1-x}$P$_x$)$_2$ and FeSe$_{1-x}$S$_x$ (with appropriate $x$ for the latter four) \cite{Niu2017, Nakajima2019, Sarkar2019, Giraldo-Gallo2018, Hayes2017, Licciardello2019}. In these systems, all related to families of topical superconductors, an intriguing interplay between the thermal and field energy scales have been established. A field-to-temperature scaling which involves a quadrature sum of the thermal and field energy scales, developed by Hayes \etal\ \cite{Hayes2017} has been successfully applied to CrAs, BaFe$_2$(As$_{1-x}$P$_x$)$_2$, FeSe$_{1-x}$S$_x$ and Ba(Fe$_{1/3}$Co$_{1/3}$Ni$_{1/3}$)$_2$As$_2$ \cite{Niu2017, Nakajima2019, Hayes2017, Licciardello2019}. However, in the hole-doped cuprate La$_{2-x}$Sr$_x$CuO$_4$, the resistivity data do not follow the quadrature scaling \cite{Giraldo-Gallo2018,Boyd2019}, while in the electron-doped cuprate La$_{2-x}$Ce$_x$CuO$_4$ ($x$=0.175), the resistivity data have been found to be proportional to the direct sum of thermal and field energies \cite{Sarkar2019}. To further understand the interplay between the magnetic field and the temperature, more examples of correlated electron systems showing LMR are needed.

Another interesting observation is that the systems discussed above are all in the vicinity of a quantum critical point,  where a $T$-linear resistivity is frequently reported \cite{Cooper2009,Bruin2013,Klintberg2012,Goh2015,Shibauchi2014,Gegenwart2008}.  Thus, the observation of LMR in these systems could hint at the emergence of a new feature associated with quantum criticality. At the quantum critical point, temperature remains the only relevant energy scale and the uncertainty principle gives $\tau\times(k_BT)\sim\hbar$. If this scattering rate ($\tau^{-1}$) dominates the charge transport the resistivity is $T$-linear. Here, $\tau^{-1}$ is simply set by fundamental constants regardless of the underlying scattering mechanism. This so-called `Planckian dissipation' has been observed in a variety of materials \cite{Zaanen2004,Cooper2009,Bruin2013, Nakajima2019, Legros2019, Cao2020}. Nevertheless, whether quantum criticality is a necessary ingredient for the observation of LMR, and its strong interplay with the $T$-linear resistivity, require further investigations.


A well-established mechanism for realizing the $T$-linear resistivity at low temperatures is to promote scattering from low-lying phonon modes \cite{Grimvall_book, Goh2015, Klintberg2012, Brown2018}. The existence of the low-lying phonon modes will  also enhance the electron-phonon coupling.
\MoGa\ is a strong electron-phonon coupling superconductor with $T_c$ of 9.8~K \cite{Verchenko2016, Verchenko2017, Sirohi2019, Marcin2019, Neha2018}, as benchmarked by the normalized specific heat jump $\Delta c_p/\gamma T_c$ and the gap-to-$T_c$ ratio $2\Delta/k_BT_c$ of 2.83 and 4.40, respectively \cite{Verchenko2016, Marcin2019}, both larger than the BCS weak-coupling values \cite{Bardeen1957a, Bardeen1957b}. Here, $\gamma$ is the Sommerfeld coefficient. Indeed, the resistivity increases linearly for $T$ between $T_c$ and $\sim$55~K, and it begins to saturate at higher $T$ \cite{Verchenko2016, Neha2018}. Thus, the $T$-linear resistivity in \MoGa\ is consistent with the strong electron-phonon coupling established from heat capacity data. In this manuscript, we report our discovery of a nonsaturating LMR in \MoGa. The $T$-linear resistivity occurs at sufficiently low temperatures where the magnetoresistance (MR) is sizeable even in a typical laboratory field, enabling a detailed investigation of the interplay between $T$-linear resistivity and LMR. Remarkably, our data exhibit a very similar behaviour to the case of La$_{2-x}$Sr$_x$CuO$_4$, despite the completely different crystal structure, Fermi surface topology and the apparent absence of quantum critical physics in \MoGa.

Single crystals of \MoGa\ were synthesized by the Ga flux method \cite{SUPP}. The electrical resistivity was measured by a standard four-terminal configuration up to 14~T in a Physical Property Measurement System by Quantum Design at CUHK, and one sample was measured up to 36~T at The National High Magnetic Field Laboratory in Tallahassee. Hydrostatic pressure was provided by moissanite anvil cells with glycerin as the pressure transmitting medium and the pressure value was obtained by ruby fluorescence at room temperature. First-principles calculations on Mo$_8$Ga$_{41}$ were performed, 
with details provided in Supplemental Material \cite{SUPP}.

 
\begin{figure}[!t]\centering
      \resizebox{8.5cm}{!}{
              \includegraphics{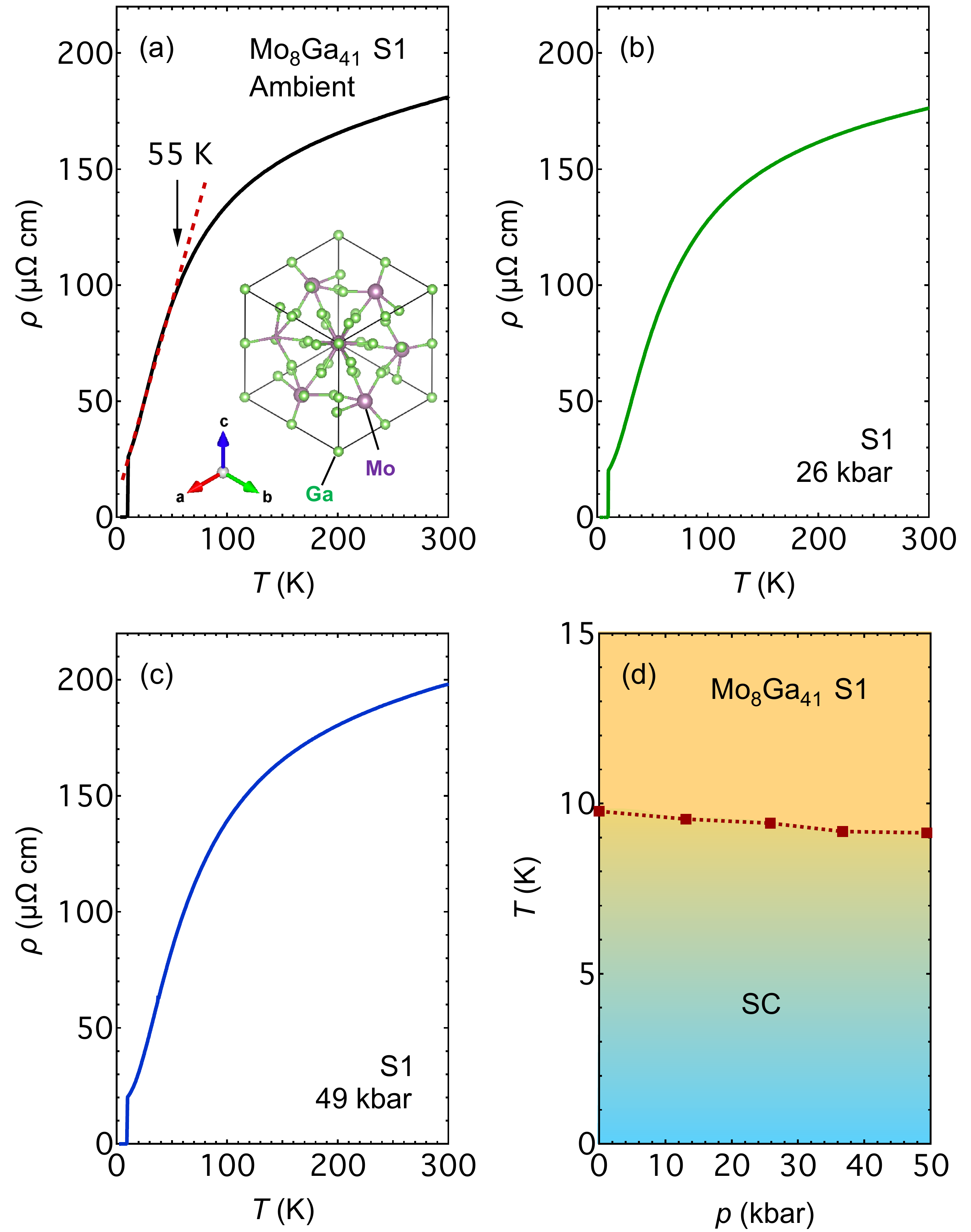}}                				
              \caption{\label{fig1}  
              (a) Temperature dependence of resistivity $\rho(T)$ in \MoGa\ (S1) at ambient pressure and zero field. The dashed line indicates the linear region. The inset shows the primitive unit cell of \MoGa\ drawn with VESTA \cite{Momma2011}. (b) $\rho(T)$ of S1 at 26~kbar. (c) $\rho(T)$ of S1 at 49~kbar. (d) Pressure dependence of $T_c$.}
\end{figure}

\begin{figure*}[!t]\centering
      \resizebox{17.0cm}{!}{
              \includegraphics{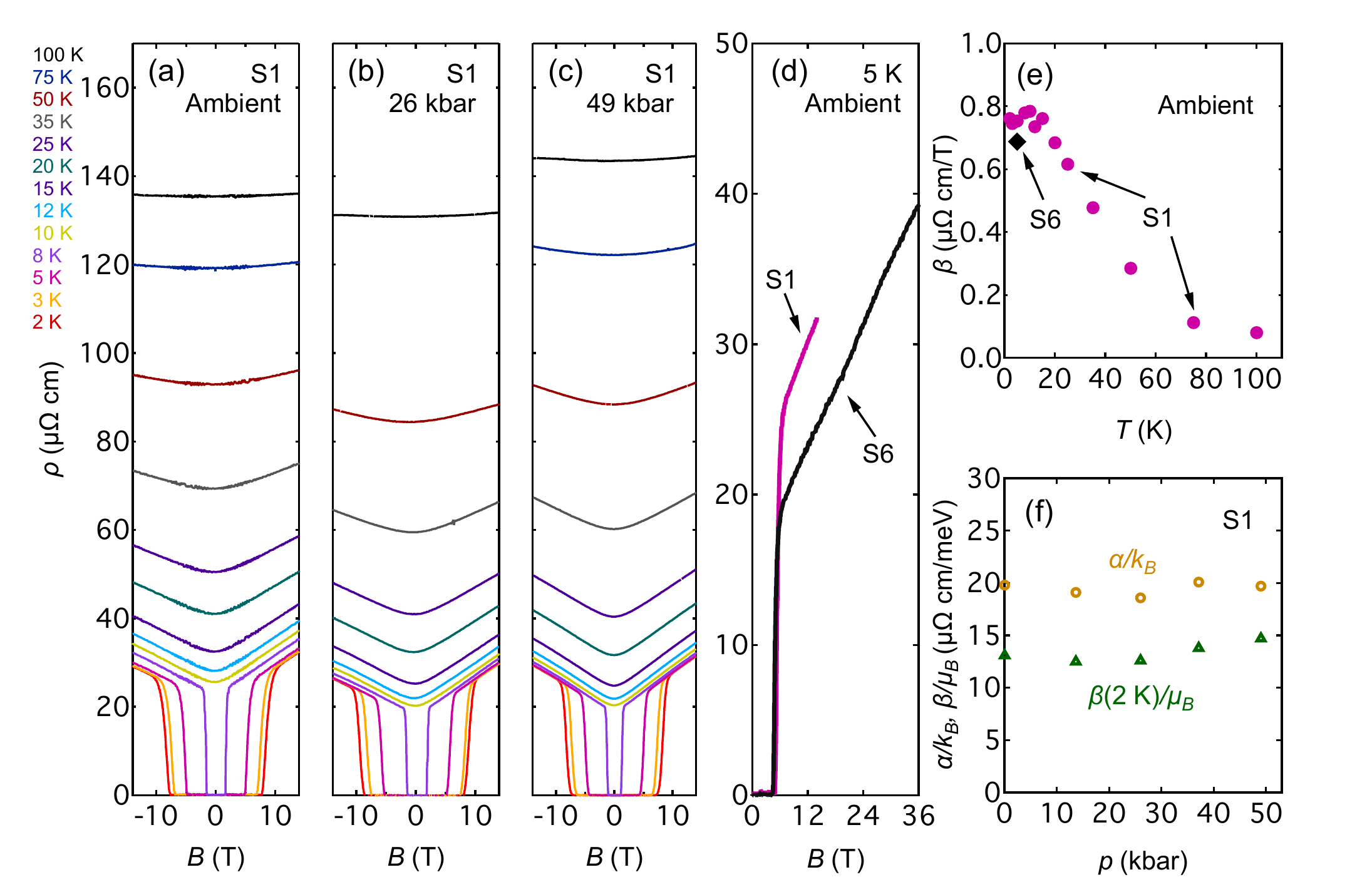}}            			
              \caption{\label{fig2} 
              Field dependence of resistivity $\rho(B)$ for S1 over a wide range of temperatures from 2~K to 100~K at (a) ambient pressure, (b) 26~kbar, and (c) 49~kbar. (d) $\rho(B)$ at 5~K up to 36~T for S6. The $\rho(B)$ trace for S1 at 5~K is included for comparison. (e) Temperature dependence of $\beta$ determined by the slope of a linear fit of $\rho_{xx}$ from 12 T to 14 T. The $\beta$ value at 5~K for S6 is included. (f) The pressure evolution of $\alpha/k_B$ (circles) and $\beta(2~{\rm K})/\mu_B$ (triangles) in energy units (${\rm \mu\Omega}$·cm/meV). 
              }            
\end{figure*}

\MoGa, which adopts the V$_8$Ga$_{41}$ structure \cite{Bezinge1984, Yvon1975}, crystalizes in a rhombohedral structure (space group $R\bar{3}$) with its primitive unit cell shown in the inset of Fig.~\ref{fig1}(a). The Mo atoms are ten-fold coordinated by Ga, forming MoGa$_{10}$ polyhedra that interconnect to form a roughly isotropic three-dimensional structure \cite{Yvon1975, Verchenko2016}. Figure~\ref{fig1}(a) shows the $T$ dependence of the electrical resistivity ($\rho$) in one of our \MoGa\ single crystals (S1) at ambient pressure. At 9.8~K, the resistivity drops sharply to zero, signaling a superconducting transition. Between $T_c$ (=9.8~K) and $\sim$55~K, $\rho(T)$ is $T$-linear with a slope $\alpha=d\rho/dT=1.71~\mu\Omega {\rm cm/K}$. In the energy unit, $\alpha/k_B=19.8~\mu\Omega {\rm cm/meV}$. With a further increase of temperature, $\rho(T)$ begins to show sign of saturation. Using an empirical `parallel resistor model' \cite{Wiesmann1977}, the observed $\rho(T)$ in \MoGa\ can be described as the effective resistivity of two parallel resistors: one has a $T$-linear resistivity from $T_c$ to 300 K and the other has a $T$-independent, saturation resistivity \cite{SUPP}. Thus, if the second resistor is not effective, $\rho(T)$ would have a large $T$-linear range as cuprates or Fe-based superconductors near the quantum critical point. Other samples exhibit similar behaviour \cite{SUPP} and these $\rho(T)$ curves are consistent with the published result \cite{Verchenko2016,Neha2018}.  Figures~\ref{fig1}(b) and \ref{fig1}(c) show $\rho(T)$ of S1 at 26~kbar and 49~kbar, respectively. The high-pressure $\rho(T)$ traces are similar to the ambient pressure curve, except for a slight nonlinearity just above $T_c$. Approximating this region as being linear, we obtain $\alpha=1.60~\mu\Omega {\rm cm/K}$ and $1.70~\mu\Omega {\rm cm/K}$ at 26~kbar and 49~kbar, respectively.  $T_c$ decreases approximately linearly with a small slope $dT_c/dp\approx-13.5~{\rm mK/kbar}$, as shown in Fig.~\ref{fig1}(d). 

We now examine the field ($B$) dependence of $\rho$ for S1. Figure~\ref{fig2}(a) plots the isothermal $\rho(B)$ at ambient pressure over a wide temperature range. $\rho(B)$ exhibits a small asymmetry upon the reversal of $B$ because of the antisymmetric Hall contribution, but is otherwise insensitive to the field direction \cite{SUPP}. The in-field data are clearly dominated by the symmetric component, which is the transverse magnetoresistance $\rho_{xx}$ and the primary interest of this work. Hence, all forthcoming analysis of the high field data have been carried out on $\rho_{xx}$. At 100~K, $\rho_{xx}(B)$ does not vary much when $B$ changes from $-14$~T to $14$~T. 
The MR, defined as $\frac{\rho_{xx}(B)-\rho_{xx}(B=0)}{\rho_{xx}(B=0)}\times100~\%$, is only 0.6~\% at 14~T.
On cooling, $\rho_{xx}(B)$ progressively becomes more sensitive to $B$. At 10~K which is just above $T_c$, $\rho_{xx}(B)$ is perfectly linear from 2.5~T to 14~T (see also Fig.~S6(b) of \cite{SUPP}) without any sign of saturation, and MR at 14~T reaches 39.8~\%. Below $T_c$, $\rho_{xx}(B)$ remains zero until the upper critical field ($B_{c2}$), above which $\rho_{xx}(B)$ grows linearly at a similar rate as the trace at 10~K. Additionally, we have conducted one ambient pressure measurement up to 36~T on \MoGa\ (S6) and found that the linear $\rho_{xx}(B)$ extends to the maximum attainable field (Fig.~\ref{fig2}(d)).
Similar magnetoresistances are also observed under pressure, with representative datasets shown in Figs.~\ref{fig2}(b) and \ref{fig2}(c). Hence, our data reveal an extraordinary magnetotransport phenomena of \MoGa: its low-temperature MR is quasilinear and nonsaturating, and LMR is robust against pressure.

The growth of the LMR on cooling can be characterized by $\beta=d\rho_{xx}/dB$. Figure~\ref{fig2}(e) displays $\beta(T)$ determined for S1 at ambient pressure using $\rho(B)$ between 12~T and 14~T. At low temperatures, $\beta$ saturates at around 0.8$~\mu\Omega {\rm cm/T}$. Such a temperature-independent $\beta$ is incompatible with a conventional scenario of an orbital MR set by the product of the cyclotron frequency $\omega_c$ and scattering time $\tau$. In the energy unit, $\beta/\mu_B=13.1~\mu\Omega {\rm cm/meV}$ at 2~K, which is comparable to $\alpha/k_B=19.8~\mu\Omega {\rm cm/meV}$ discussed earlier. The pressure dependences of $\alpha/k_B$ and $\beta(2~{\rm K})/\mu_B$ for S1 are summarized in Fig.~\ref{fig2}(f). Our central finding here is that the magnetic field is as efficient as temperature in driving the linear increase in the resistivity, hinting at the equivalence of field energy and thermal energy in controlling the scattering rate.

The LMR discovered in \MoGa\ resembles the scale-invariant MR in La$_{2-x}$Sr$_x$CuO$_4$ even at the visual level. In the latter system with hole doping level $p$=0.190, $\beta/\mu_B$ saturates at low temperature with a value $5.2~\mu\Omega {\rm cm/meV}$, while $\alpha/k_B=11.8~\mu\Omega {\rm cm/meV}$ \cite{Giraldo-Gallo2018}. These values are comparable to the case of \MoGa. Furthermore, $(\alpha/k_B)/(\beta/\mu_B)$ is also similar for both systems: the ratio is 2.3 for La$_{2-x}$Sr$_x$CuO$_4$ ($p$=0.190), and 1.5 for \MoGa\ (S1) at ambient pressure. These similarities are surprising, given that the two systems have very distinct character: the crystal and the electronic structures of \MoGa\ are significantly more three-dimensional compared with La$_{2-x}$Sr$_x$CuO$_4$, and the Fermi surface of \MoGa\ is also more complicated with multiple sheets.

Experiments on other \MoGa\ samples at ambient pressure give $\alpha/k_B=16.2, 22.0, 24.5, 14.8$ and $17.2~\mu\Omega {\rm cm/meV}$ for S2--S6 respectively \cite{SUPP}. Interestingly, although $\alpha/k_B$ shows a standard deviation of 19\% around the mean value ($19.1~\mu\Omega {\rm cm/meV}$) across the six samples, $(\alpha/k_B)/(\beta/\mu_B)$ exhibit a much smaller distribution: the ratio is 1.5, 1.5, 1.6, 1.7, 1.5 and 1.4 for S1--S6 respectively. This reinforces our observation that the magnetic field and the temperature are similarly efficient in driving the linear increase in the resistivity.
\begin{figure}[!t]\centering
       \resizebox{8.5cm}{!}{
              \includegraphics{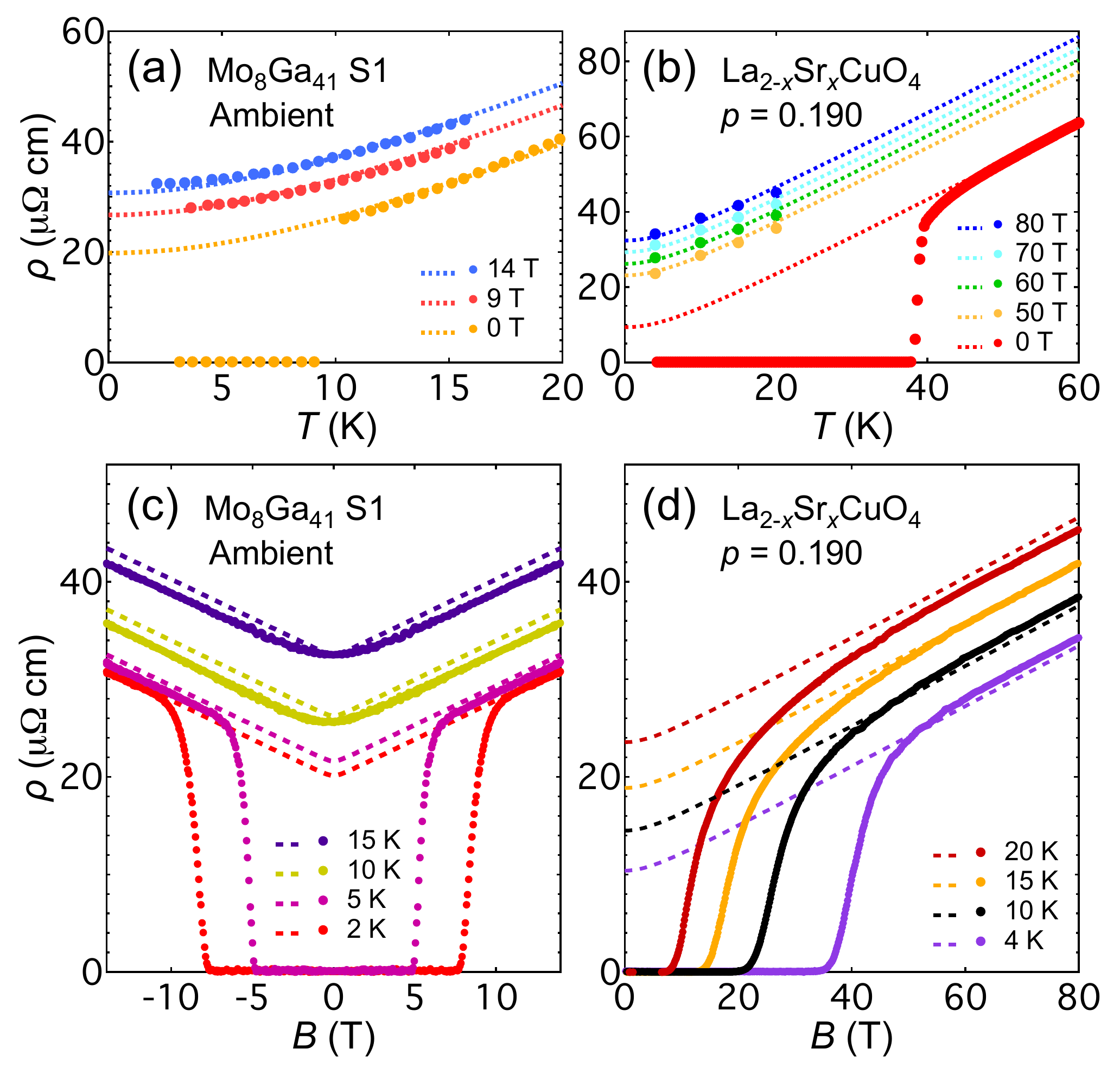}}                				
              \caption{\label{BTfig}
              $\rho(T)$ at fixed $B$ for (a) \MoGa\ (S1) at ambient pressure and (b) La$_{2-x}$Sr$_x$CuO$_4$ ($p$=0.190). The low-temperature isothermal $\rho(B)$ of (c) \MoGa\ (S1) at ambient pressure and (d) La$_{2-x}$Sr$_x$CuO$_4$. The data of La$_{2-x}$Sr$_x$CuO$_4$ come from Ref.~\cite{Giraldo-Gallo2018}. For this figure, the open symbols are experimental data while the dashed curves are simulations based on Eqn.~\ref{BTeqn}. For \MoGa\ (S1), $\alpha$=1.71~$\mu\Omega {\rm cm/K}$, $\beta$=0.8~$\mu\Omega {\rm cm/T}$, $\rho_T$=19.5$\pm$0.1~$\mu\Omega {\rm cm}$ and $\rho_B$=0.30$\pm$0.14~$\mu\Omega {\rm cm}$. For La$_{2-x}$Sr$_x$CuO$_4$ ($p$=0.190), $\alpha$=1.02~$\mu\Omega {\rm cm/K}$, $\beta$=0.31~$\mu\Omega {\rm cm/T}$, $\rho_T$=7.5$\pm$0.4~$\mu\Omega {\rm cm}$, $\rho_B$=1.82$\pm$0.03~$\mu\Omega {\rm cm}$.  
              }
\end{figure}

To further understand the interplay between the temperature and the
magnetic field, we have analyzed our data with the scaling
proposed by
Hayes \etal\ for BaFe$_2$(As$_{1-x}$P$_x$)$_2$ \cite{Hayes2017}:
$\rho(B,T)=\rho(0,0)+\sqrt{(\alpha T)^2+(\beta B)^2}$, where $\alpha$
and $\beta$ are constants. Our data cannot be described by this
quadrature sum, even at low temperatures when $\beta$ is insensitive
to temperature \cite{SUPP}. That is because at a
given $T_{\textrm{fix}}$, Hayes' scaling predicts that the linear-in-$B$
behavior only appears when 
$B \gg \alpha T_{\textrm{fix}}/\beta$. However, in
Mo$_8$Ga$_{41}$, LMR can be found even when
$B \ll \alpha T_{\textrm{fix}}/\beta$ \cite{footnote}. 
Similarly, Hayes' scaling also
fails for La$_{2-x}$Sr$_x$CuO$_4$ because at a
given magnetic field $B_{\textrm{fix}}$, a linear-in-$T$ resistivity has been found to
persist to a low temperature much smaller than $\beta B_{\textrm{fix}}/\alpha$~\cite{Giraldo-Gallo2018}.

Instead of Hayes' scaling, we find that our low temperature
data can be adequately captured by the following empirical formula:
\begin{equation}
\rho(B,T) = \sqrt{\rho_T^2+(\alpha T)^2} + \sqrt{\rho_B^2+(\beta B)^2}.
\label{BTeqn}
\end{equation}
Because of the relatively low $T_c$, the low-temperature normal state
of \MoGa\ can be fully exposed with a sufficiently high laboratory
field. At 14~T and 9~T, we can access the normal state of \MoGa\ down
to 2.0~K (our lowest temperature) and 3.7~K, respectively, giving an
opportunity to test Eqn.~\ref{BTeqn}. Because $\beta$ begins to show
temperature dependence above $\sim$20~K, we restrict our analysis to
data below 15~K. To avoid introducing four free parameters, both
$\alpha$ and $\beta$ are fixed to values determined earlier for \MoGa\ (S1):
$\alpha=1.71~\mu\Omega {\rm cm/K}$, $\beta=0.8~\mu\Omega {\rm
  cm/T}$. The parameters $\rho_T$ and $\rho_B$ are determined self-consistently using $\rho(14~{\rm T}, T)$, 
$\rho(9~{\rm T}, T)$ and $\rho(0~{\rm T}, T\!\in\![T_c,20~K])$~\cite{footnote2}. With $\rho_T$, $\rho_B$, $\alpha$ and $\beta$
determined, we can then compare our scaling formula with the experimental data
for any combination of $B$ and $T$: the curves simulated with our scaling formula (dashed curves) agree nicely with the experimental normal state data (open symbols), as displayed in Figs.~\ref{BTfig}(a) and \ref{BTfig}(c).

We now examine La$_{2-x}$Sr$_x$CuO$_4$~($p$=0.190), the other system that defies Hayes' scaling \cite{Giraldo-Gallo2018}. Similar to \MoGa, we only analyze the magnetotransport data below $\sim$20~K, where $\beta$ is a constant. Following the identical procedure, we keep $\alpha$ and $\beta$ constant, and use $\rho(B_{\rm fix},T)$ at $B_{\rm fix}$=50, 60, 70 and 80~T together with $\rho(0~{\rm T}, T\!\in\![50~{\rm K},60~{\rm K}])$~\cite{footnote2} to determine $\rho_T$ and $\rho_B$ self-consistently \cite{SUPP}. With the values of $\rho_T$ and $\rho_B$ thus obtained, we simulate $\rho(B,T)$, as displayed in Figs.~\ref{BTfig}(b) and \ref{BTfig}(d). Our scaling formula successfully describes the normal state of La$_{2-x}$Sr$_x$CuO$_4$ too.

Our empirical model shows that at a fixed temperature $T_{\rm fix}$, $\rho(B,T_{\rm fix})$ approaches the zero field limit quadratically. 
This weak-field behaviour is commonly seen in many systems \cite{Pippard_book}. Similarly, for a fixed field $B_{\rm fix}$, the model also predicts a quadratic $\rho(B_{\rm fix},T)$ at the zero temperature limit. In particular, such a behaviour is guaranteed for $B_{\rm fix}=0$. Thus, the zero field resistivity turns from linear at moderate temperatures to quadratic at the lowest temperature. Our scaling formula describes the zero-field $\rho(T)$ of both \MoGa\ and La$_{2-x}$Sr$_x$CuO$_4$ well (see Fig.~S5 of \cite{SUPP}). In \MoGa, we further note that in the standard framework of electron-phonon scattering, the linear-in-$T$ resistivity only kicks in when $k_BT$ is greater than some characteristic energy of the phonon modes \cite{Grimvall_book}. Otherwise, a higher temperature exponent is expected. We argue that in \MoGa, the characteristic phonon energy is lowered because of an abundance of low-lying phonon modes at finite wavevectors, but this characteristic phonon energy remains finite. At sufficiently low temperature, the linear-in-$T$ channel is not yet active, but the usual $T^2$ dependence due to electron-electron interaction dominates the low temperature part of the data.

Although the central aims of this paper are to report the discovery that $(\alpha/k_B)\sim(\beta/\mu_B)$ and to present the new empirical scaling, we close the paper by a brief comment on the applicability of two popular mechanisms of nonsaturating LMR. The first scenario involves the quantum magnetoresistance when a given Fermi surface sheet reaches the extreme quantum limit \cite{Abrikosov1998, Abrikosov2000}. If this Fermi surface sheet dominates the magnetotransport, a nonsaturating LMR can be observed \cite{Niu2017}. However, this scenario is challenging for \MoGa\ with complicated, multiple Fermi surface sheets \cite{SUPP}. Although DFT calculations
show that within some parameter range, a small electron pocket with linear dispersion can
appear around the $Q$ point of Brillouin zone and thus the highly mobile electrons in the
pocket can potentially be driven into the extreme quantum limit, it is difficult to neglect the contributions from other Fermi surface sheets. Thus, quantum linear magnetoresistance is unlikely to be the sole explanation. The second scenario is related to disorder of the system, which can also result in nonsaturating LMR \cite{Parish2003, Patel2018, Singleton2020}. To explore this scenario, we measured the MR of vanadium-doped \MoGa, as presented in Supplemental Material \cite{SUPP}.
The ratio $\rho(300~{\rm K})/\rho(10~{\rm K})$ can be taken as an indicator of sample purity. Although $\rho(300~{\rm K})/\rho(10~{\rm K})$ of \MoVGa\ is about 3 -- 4 times lower than a typical \MoGa\,
the MR remains nonsaturating and linear in both cases. Thus, disorder-induced LMR is also not expected to play a dominant role. The underlying mechanism for LMR in \MoGa\ remains a topic for future investigations. Such a mechanism would also need to explain the interplay between LMR and the $T$-linear resistivity.

In summary, we have conducted a comprehensive measurement of the tranverse magnetoresistance in \MoGa. 
We discover a robust nonsaturating linear magnetoresistance that persists under pressure up to at least 49~kbar, and in magnetic field up to at least 36~T. An interesting interplay between the linear magnetoresistance and the $T$-linear resistivity -- similar to the observation in optimally doped La$_{2-x}$Sr$_x$CuO$_4$ -- is revealed, which establishes that the temperature and magnetic field are equally capable of driving the linear increase of the resistivity, as illustrated by our finding that $(\alpha/k_B)\sim(\beta/\mu_B)$. A new empirical model is developed to describe the low-temperature $\rho(B,T)$. 
The linear magnetoresistance, and the similarity between $(\alpha/k_B)$ and $(\beta/\mu_B)$ are also established in \MoVGa, which is more disordered than \MoGa. With the apparent absence of quantum critical physics, \MoGa\ is less strange than a typical ``strange metal" phase, and thus the data presented here can be a useful reference for the eventual understanding of ``strange metal" physics. 

\begin{acknowledgments}

We acknowledge Esteban Paredes Aulestia, Stanley W. Tozer, Scott A. Maier and Bobby Joe Pullum for experimental support and Xiaofang Zhai, Wing Chi Yu and Kai Liu for discussions. The work was supported by Research Grants Council of Hong Kong (CUHK 14300418, CUHK 14300117, CUHK 14300419), CUHK Direct Grant (4053345, 4053299), the Ministry of Science and Technology of Taiwan (MOST-106-2112-M-006-013-MY3), the National Natural Science Foundation of China (11774236), and the NYU University Research Challenge Fund. H. C. and Y. F. acknowledge the computational resources provided by NYU HPC resources at New York, Abu Dhabi and Shanghai campuses. A portion of this work was performed at the National High Magnetic Field Laboratory, which is supported by National Science Foundation Cooperative Agreement No. DMR-1644779, the State of Florida, and the U.S. Department of Energy.

\end{acknowledgments}


\begin{thebibliography}{50}
\bibitem [{foo()}]{footnote2}%
  \BibitemOpen
  \href@noop {} {}\bibinfo {note} {The inclusion of the zero-field $\rho(T)$ results in a better estimate of $\rho_B$. At zero field $\beta$ is not relevant. Hence data at higher temperatures where $\beta$ is not a constant can be included. For more information, see Section VI of~\cite{SUPP}.}
  \BibitemShut {Stop}%
\end{thebibliography}

\providecommand{\noopsort}[1]{}\providecommand{\singleletter}[1]{#1}%

\end{document}